\documentclass{article}
\usepackage{authblk}
\usepackage{amsmath}
\usepackage[utf8]{inputenc}
\usepackage{caption}
\usepackage{graphicx}
\graphicspath{ {./figs/} }

\begin{document}
\title{High-performance orbit-following code ASCOT5 for Monte Carlo simulations in fusion plasmas}
\author[1]{J. Varje}
\author[1]{K. Särkimäki}
\author[1]{J. Kontula}
\author[1]{P. Ollus}
\author[1]{T. Kurki-Suonio}
\author[1]{A. Snicker}
\author[1]{E. Hirvijoki}
\author[1,2]{S. Äkäslompolo}
\affil[1]{Department of Applied Physics, Aalto University, P.O. Box 11100, 00076 AALTO, Finland}
\affil[2]{Max-Planck-Institut für Plasmaphysik Teilinstitut Greifswald, Wendelsteinstr. 1, 17491 Greifswald, Germany}
\date{\today}
\maketitle
\section*{Abstract}
We present a novel implementation of a Monte Carlo particle-following code for solving the distribution function of minority species in fusion plasmas, called ASCOT5, and verify it using theoretical results for neoclassical transport. The code has been developed from ground up with an OpenMP-MPI hybrid paradigm to take full advantage of current and next generation many-core CPUs with multithreading and SIMD operations. Up to 6-fold increase in performance is demonstrated compared to a previous version of the code which only utilizes MPI. The physics model of the code is comprehensively validated against existing theoretical work, and it is shown to faithfully reproduce neoclassical diffusion across three different collisionality regimes. In simulations for realistic tokamak plasmas, including complex non-axisymmetric geometry, ASCOT5 is verified to reproduce results from the previous version ASCOT4.

\section{Introduction}
As experimental fusion research progresses towards reactor era with larger, more powerful and more complex devices such as Wendelstein 7-X, JT-60SA and ITER, increasing demand is placed on simulation results in support of design, preparation and operation of these experiments. In particular, predicting the behavior of fast ions in various plasma conditions is vital, both for achieving high performance and ensuring machine protection in terms of fast ion wall loads. Inclusion of additional physical processes affecting fast ion confinement, such as 3D magnetic perturbations, time-dependent MHD perturbations and complex synergies between plasma heating methods, further increases the complexity of the simulations.

At the same time, modern supercomputers are being designed for an ever-increasing degree of parallel processing at multiple levels. Typically, it has been sufficient to divide the problem into multiple simultaneously executing programs, which communicate with each other over a network using techniques such as MPI. However, modern processors have built-in parallelism in the form of multithreading as well as SIMD vector operations, which effectively perform the same arithmetical operations simultaneously for multiple values. Taking full advantage of the performance of these features requires modern programming methods and a code developed with the hardware features in mind.

A common computational approach to problems where finite orbit width is important, or the geometry is non-trivial, is charged particle orbit-following, where the trajectories of charged particles in a plasma are solved. This approach can further be used in a Monte Carlo method to solve the distribution of the particles in complex geometries, where many other approaches become inefficient. ASCOT is one such code, designed to follow minority particles, such as fast ions, in a fusion plasma in a realistic three-dimensional geometry. Up until ASCOT4, the previous versions of the code had been parallelized up to single core level with basic MPI parallelization. This approach has recently been found lacking on modern multithreaded, SIMD-enabled CPUs, in particular on many-core platforms such as Xeon Phi. Lack of multithreading and poor vectorization performance, as well as limited memory due to lack of shared memory programming, were identified as significant bottlenecks.

To fix these issues, a full rewrite of the codebase was undertaken, resulting in ASCOT5. The simulation code was rewritten in C using a hybrid OpenMP + MPI approach. The core simulation loop was explicitly written with SIMD execution in mind. The code was organized in a highly modular structure, with clear interfaces between different modules, in an attempt to improve the long-term maintainability of the code, a common problem with scientific codes. 

In this paper we describe the ASCOT5 code and the particular solutions developed to utilize modern multithreaded, SIMD-enabled CPUs. Section~\ref{sec:physmodel} describes the core physics model, and section~\ref{sec:implementation} the parallelized code implementation. In section~\ref{sec:verification}, the correctness of the code is verified both against theoretical results as well as results with an older version of the code. Finally, the performance of the hybrid OpenMP + MPI code is compared to the earlier MPI-only version, and the scalability of the code is demonstrated.

\section{Physics model}
\label{sec:physmodel}

Like its predecessors~\cite{hirvijoki2014ascot, hirvijoki2015monte}, ASCOT5 solves the distribution function of a minority species in a toroidal magnetic confinement fusion device. 
The evolution of the distribution function $f_a(\mathbf{x}, \mathbf{v}, t)$ for a test particle species $a$ is described by the Fokker-Planck equation
\begin{equation}
\frac{\partial f_a}{\partial t}+\mathbf{v} \cdot \nabla f_a + \frac{q_a}{m_a} ( \mathbf{E} + \mathbf{v} \times \mathbf{B}) \cdot \nabla_{\mathbf{v}} f_a = 
\sum_{b} -\nabla_{\mathbf{v}}\cdot\left[\mathbf{a}_{ab}f_a -\nabla_{\mathbf{v}}\cdot (\mathbf{D}_{ab}f_a)\right]
\end{equation}
where $\mathbf{x}$, $\mathbf{v}$ and $t$ are the position, velocity and time coordinates, $q_a$ and $m_a$ are test particle charge and mass, $\mathbf{E}$ is the electric field and $\mathbf{B}$ is the magnetic field.
Coulomb collisions between the test particles and the background plasma species, which are assumed to be Maxwellian, are included via the friction coefficient $\mathbf{a}_{ab}$ and the diffusion coefficient $\mathbf{D}_{ab}$.

ASCOT5 finds the approximate solution to the Fokker-Planck equation by solving the corresponding Langevin equation for a large number of markers that represent the distribution function $f_a$.
The Langevin equation, i.e., the equation of motion for each marker, is
\begin{equation}
d\mathbf{z} = \left[\mathbf{\dot{z}} + \mathbf{a}(\mathbf{z},t)\right] dt + \boldsymbol{\sigma}(\mathbf{z},t)\cdot d\boldsymbol{\mathcal{W}},
\end{equation}
where $\mathbf{z}$ are the phase space coordinates $\mathbf{x}$ and $\mathbf{v}$, $\mathbf{\dot{z}}$ is the coordinate time-derivative without Coulomb collisions, $\mathbf{a} = \sum_b\mathbf{a}_{ab}$, $\boldsymbol{\sigma}$ is defined by the relation $\sum_b 2\mathbf{D}_{ab} = \boldsymbol{\sigma}\boldsymbol{\sigma}^T$, and $\boldsymbol{\mathcal{W}}$ is a 6D vector of independent Wiener processes.
The distribution function $f_a$ is approximated by discretizing the equation in time, integrating it numerically, and at each timestep summing marker weight to a correct bin in a $\mathbf{x}$, $\mathbf{v}$ histogram.

ASCOT5 can also solve the guiding center distribution function.
The guiding center equations of motion are based on the non-canonical Hamiltonian dynamics~\cite{cary2009hamiltonian}, with the guiding center coordinates being $\mathbf{X}$, $v_\parallel$ and $\mu$ (guiding center position, parallel velocity, and magnetic moment).
The equations of motion and the transformation from particle to guiding center phase space are done to first order, and are valid in the relativistic regime \cite{hirvijoki2015guiding}.
However, the guiding center collision operator is a zeroth order operator adapted from~\cite{hirvijoki2013monte}.

For numerical integration, the Hamiltonian part of the Langevin equation, i.e. $\mathbf{\dot{z}}$ term, is solved first at the beginning of a timestep and the coordinates updated accordingly, and then the collisional part is solved and coordinates are updated.
The benefit of this separation is that one may use different numerical scheme to solve the Hamiltonian motion, i.e. solving an ordinary differential equation, than what is used to solve the collisional part, which is a stochastical differential equation.
For example, for fast particles it is usually the Hamiltonian motion that dominates the marker motion which can be solved efficiently using a higher-order Runge-Kutta method, whereas higher order stochastic Runge-Kutta methods may not offer any benefit in terms of computational cost.

The gyro orbit is solved numerically with a fixed timestep using the energy-conserving, relativistic volume-preserving algorithm~\cite{zhang2015volume}. 
For the guiding center, the equations of motion are solved either with fixed timestep using RK4, or with adaptive timestep using the Cash-Karp algorithm, where the error is estimated using the difference between the fourth and fifth-order solutions. 
The Coulomb collisions are solved using the numerical schemes discussed in~\cite{sarkimaki2018adaptive}, with small adaptions.
Euler-Maruyama method is used in particle and fixed timestep guiding center modes, and Milstein method is used with the adaptive guiding center mode.

For a steady-state distribution function, the markers are simulated from their source location until they reach certain criteria for their end condition. For example, in the case of neutral beam injection (NBI) heating of a tokamak plasma, the orbits of the markers are followed as they slow down. Once the markers approach the temperature of the thermal plasma, defined as a multiple of the local temperature, they are considered thermalized and the simulation is terminated.

Particle losses are evaluated by checking for intersections between the particle trajectory and a surface representing the first wall of the device. The wall consists either of a 2D poloidal contour of the limiting surfaces, or an arbitrary 3D triangle mesh. The 3D mesh is stored in an octree structure to enable efficient searching of intersecting triangles, while limiting the number of memory accesses.

ASCOT5 requires magnetic field, electric field, plasma profiles, and wall model to be given as an input.
Inputs are implemented via interfaces, meaning that ASCOT5 is not restricted to a specific method when e.g. evaluating magnetic field at marker position.
For example, magnetic field can be either evaluated analytically using a parametrized model for the poloidal flux~\cite{cerfon2010one}, or numerically interpolated with either 2D (axisymmetric field) or 3D splines (3D field).
Different implementations for the inputs are added as the need arises.

\section{Implementation}
\label{sec:implementation}

ASCOT5 was designed from ground up to maximize utilization of the SIMD vector operations. All calculations in the physics simulation loop operate on arrays of length $N_\mathrm{SIMD}$, which can be set depending on the platform. For processors with 256-bit AVX2 operations (Haswell etc.) this is equivalent to four 64-bit double precision floats, while for 512-bit AVX512 processors (Knights Landing, Skylake etc.) arrays of eight doubles are used.

All loops during particle simulation are parallelized with OpenMP \texttt{simd} pragmas. Additionally, all functions called during simulation, such as 3D magnetic field data access and interpolation, are implemented as vectorized functions declared using \texttt{declare simd} pragmas. This enables the compiler to execute nearly the entire core simulation loop with vector operations.

The physics calculations are identical for all particles being followed, so efficient lockstep evaluation of SIMD operations is always possible. However, as the simulation duration for each particle is dependent on their particular trajectory, not all elements are simulated for the same duration. A swapping mechanism was implemented where, after each iteration, particles that have reached their end condition are stored in an array for completed particles, and a fresh particle is retrieved from a queue to continue simulation in the particular slot in the $N_\mathrm{SIMD}$ arrays.

To enable multithreading, a number of worker threads, each operating on a single set of $N_\mathrm{SIMD}$ arrays, are launched and allowed to perform their simulation independently. Concurrent access is needed for tallying the particle distributions in the histograms, which is implemented using \texttt{atomic} pragmas. Swapping particles is also protected by \texttt{critical} sections to avoid race conditions. In addition to the simulation threads, additional maintenance threads are used for monitoring and reporting simulation progress.

Finally, MPI is used to launch multiple processes across a number of nodes in a typical supercomputing environment. During the simulation, each process is independent, storing its results only at the end, to be combined after the completion of the simulation.

\section{Verification}
\label{sec:verification}

Thorough verification of the code has been performed, first by verifying the code against known theoretical results~\cite{wesson2011tokamaks,helander2005collisional} for charged particles undergoing classical and neoclassical transport, and then benchmarking simulation results in realistic cases with ASCOT4, which in turn has been extensively benchmarked and validated both with other similar codes \cite{kramer2013simulation,asunta2015} as well as experimental measurements \cite{akaslompolo2016,varje2017,siren2017,varje2019,siren2019}.

\subsection{Comparison to theoretical results}

The particle transport in a uniform magnetic field, i.e. classical transport, is diffusive with a diffusion coefficient
\begin{equation}
D_C = \frac{1}{2}\rho_g^2\nu_{ab}
\end{equation}
where $\rho_g= mv_\perp/|q|B$ is the test particle Larmor radius and $\nu_{ab}$ collision frequency.
The collision frequency for electron-ion collisions is
\begin{equation}
\nu_{ei} = \sqrt{\frac{2}{\pi}}\frac{e^2n \ln\Lambda}{12\pi\varepsilon_0^2\sqrt{m_e T^3 }},
\end{equation}
and for ion-electron collisions $\nu_{ie} = (m_e/m_i)\nu_{ei}$.
Here $n$ is the density, $T$ the temperature, $\varepsilon_0$ the vacuum permittivity, $e$ the elementary charge, and $\ln\Lambda$ the Coulomb logarithm.

The transport in a tokamak, i.e. neoclassical transport, can be divided into three regimes depending on the collisionality, $\nu^* = \nu_{ab}/\Omega_\mathrm{orb}$, which is the ratio of the collision frequency and the test particle orbit transit frequency $\Omega_\mathrm{orb}= v\sqrt{\epsilon}/(\sqrt{2}q_pr)$, where $q_p$ is the safety factor, $\epsilon=r/R$, is inverse aspect ratio, $r$ is the minor radius and $R$ is the major radius.
At the low-collisional \emph{banana regime}, $\nu^*\ll \epsilon^{3/2}$, the diffusion coefficient is
\begin{equation}
D_B = \epsilon^{-3/2}q_p^2 D_C.
\end{equation}
At the opposite high-collisional limit, $\nu^* \gg 1$, is the \emph{Pfirsch-Schl\"uter regime} where the diffusion coefficient is
\begin{equation}
D_{PS} =  q_p^2D_C,
\end{equation}
In between of these two regimes lies the \emph{plateau regime}, where the diffusion coefficient,
\begin{equation}
D_P = q_p^2 \frac{D_C}{\nu^*},
\end{equation}
is constant.

\begin{figure}[!t]
\centering
\includegraphics[width=0.9\textwidth]{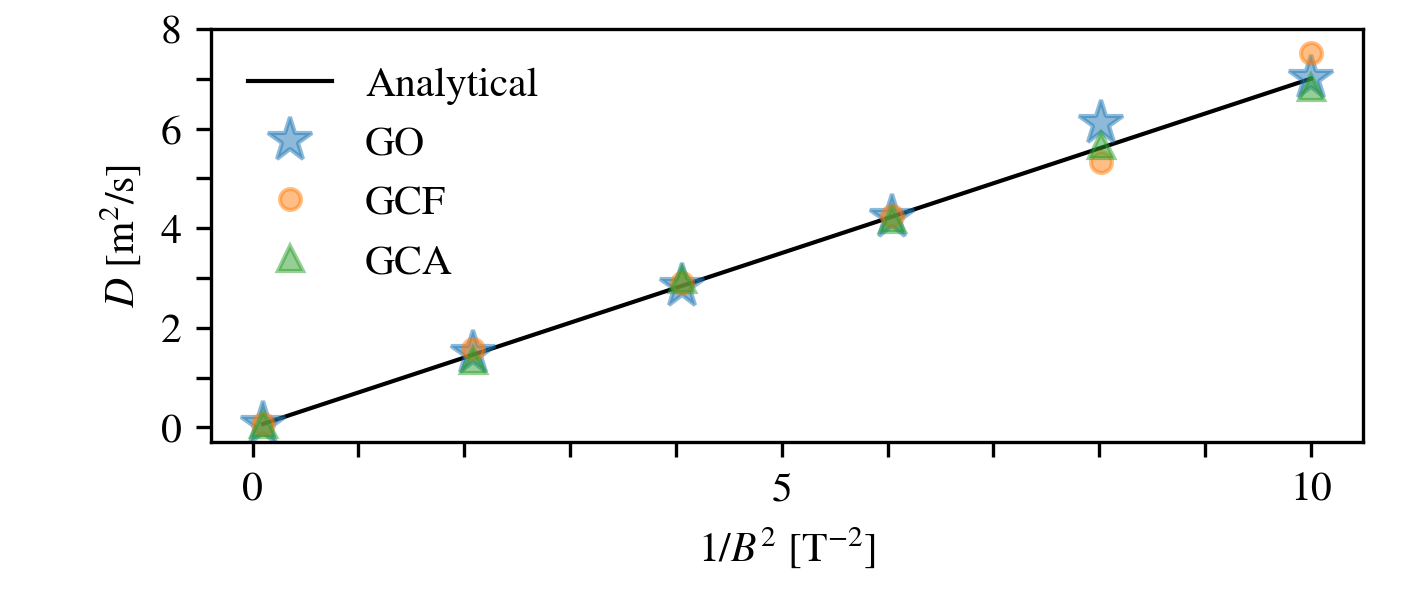}
\caption{Classical diffusion coefficient. The analytically calculated numbers are given by the solid line, while the various symbols correspond to values obtained with ASCOT5 using various approaches to orbit following as explained in the text.}
\label{fig:testclassical}
\end{figure}

Reproducing classical transport with ASCOT5 is straightforward.
Markers are launched from a same initial position and simulated for a few collision times.
After the simulation, the variance calculated from the change in markers' position can be used to calculate the diffusion coefficient as $D=\mathrm{Var}[(\Delta x)]/ 2\Delta t$.
Here 200 markers are used.
Figure~\ref{fig:testclassical} shows the $1/B^2$ scaling for test protons in an uniform field and in electron-proton plasma, when only the ion-electron collisions ($n=10^{20}$ m$^{-3}$, $T=10$ keV) are considered (for simplicity).
The abbreviations in the figure stand for the different simulation modes: Gyro-Orbit simulation, Guiding Center Fixed time-step simulation, and Guiding Center Adaptive time-step simulation.

\begin{figure}[!t]
\centering
\includegraphics[width=0.9\textwidth]{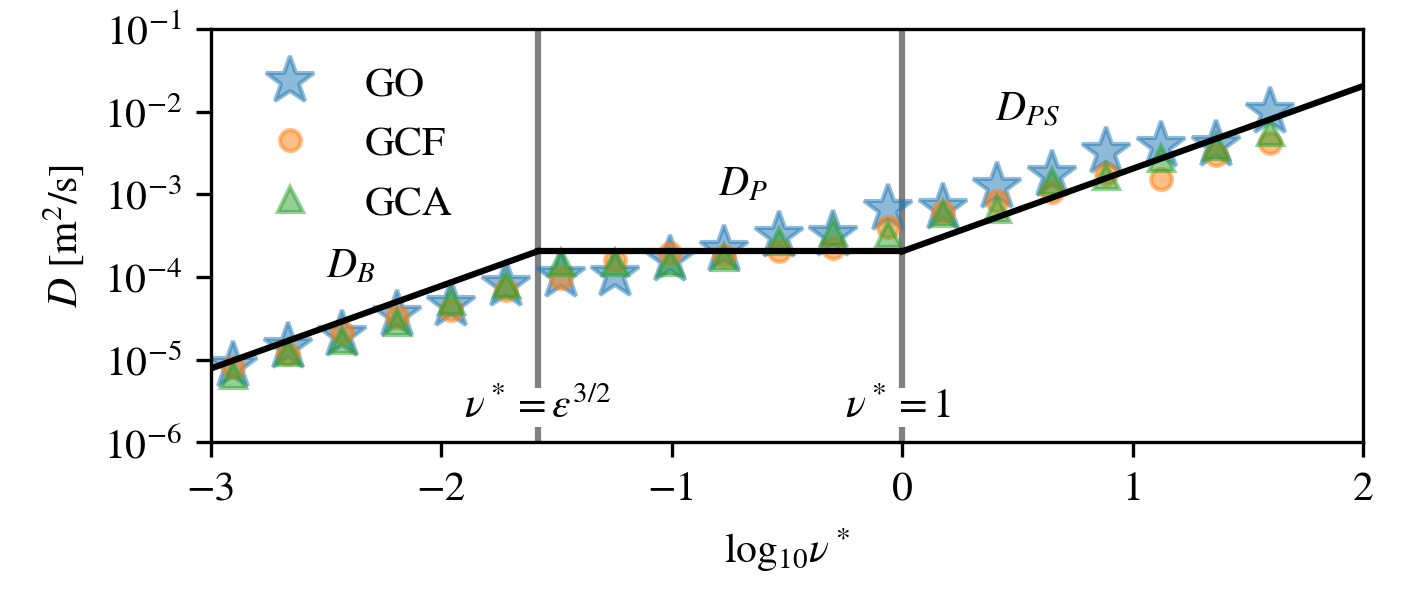}
\caption{Neoclassical diffusion coefficient across the different collisionality regimes (note the logarithmic scales). The analytically calculated results for the circular geometry are given by the solid line, while the various symbols have the same meaning as in Fig.\ref{fig:testclassical}}.
\label{fig:testneoclassical}
\end{figure}

The formulas for neoclassical transport are accurate only for a circular plasma, which we take into account by choosing the parameters for the analytical magnetic field accordingly (with $R=6.2$ m, $r=0.6$ m, $B_0=5.3$).
For the neoclassical transport test we model test electrons and consider only electron-ion collisions.
The collisionality is adjusted by varying density ($n=10^{17}$~--~$10^{22}$ m$^{-3}$, $T=1$ keV) while keeping other parameters constant.
The diffusion coefficient is again evaluated from the change in markers' position.
A total of 100 markers are simulated and they are launched from the same radial position.
Figure~\ref{fig:testneoclassical} confirms that ASCOT5 produces correct transport in all regimes.

\begin{figure}[th]
\centering
\includegraphics[width=0.5\textwidth]{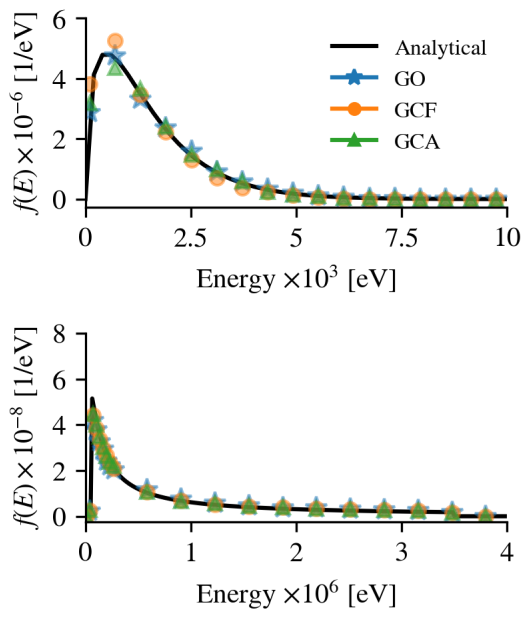}
\caption{Energy distribution of fusion alpha particles corresponding to thermal equilibrium (top) and slowing-down (bottom) distributions given by analytical theory (solid line) and ASCOT5 simulations (symbols).
}
\label{fig:slowingdown}
\end{figure}

Common application of ASCOT5 is to study neoclassical transport of fast ions during their slowing-down process.
Therefore, we verify that ASCOT5 produces correct slowing-down energy distribution,
\begin{equation}
f_s(E) =  \frac{S}{2\nu_s E}\frac{H(E_0-E)H(E-\beta E_\mathrm{th,i})}{1+(E_\mathrm{crit}/E)^{3/2}},,
\end{equation}
where $S$ is the birth rate of fast ions, $H$ is Heaviside step function and
\begin{equation}
\nu_s = \frac{6\sqrt{2\pi^2T^3}\varepsilon_0^2 m_a}{Z_\alpha^2 e^4 m_e^{\frac{1}{2}}n \log\Lambda}
\end{equation}
is the slowing down rate with $m_a$ and $Z_a$ being test particle mass and charge number, respectively.
Test particle initial energy is $E_0$, $E_\mathrm{min}$ is the energy below which a particle is considered thermalized, and
\begin{equation}
E_{crit} = \frac{m_aT}{m_i} \left( \frac{3\sqrt{\pi}m_e}{4m_i}\right)^\frac{2}{3},
\end{equation}
is the critical energy below which scattering with background ions starts to dominate.
Once particles have slowed-down, they become part of the thermal bulk whose energy distribution is Maxwellian 
\begin{equation}
f_M(E) = 2 \sqrt{\frac{E}{\pi}} T^{-3/2} e^{-E/T}
\end{equation}
Figure~\ref{fig:slowingdown} verifies that both distributions are produced when fusion alpha particles, with initial energy $E_0=3.5$ MeV, are simulated and energy distribution collected. 
Particles are simulated until they slow down to $E_\mathrm{min}= 50 T$. 
Equilibrium distribution is collected by initializing markers with $E=T$ and simulating them for a few collision times.

Additionally, it was verified that, with Coulomb collisions disabled, the energy, magnetic moment, and canonical momentum are conserved to a good accuracy.

\subsection{Fast ion simulations in 2D and 3D}

\begin{figure}[ht]
\centering
\includegraphics[height=0.45\textwidth]{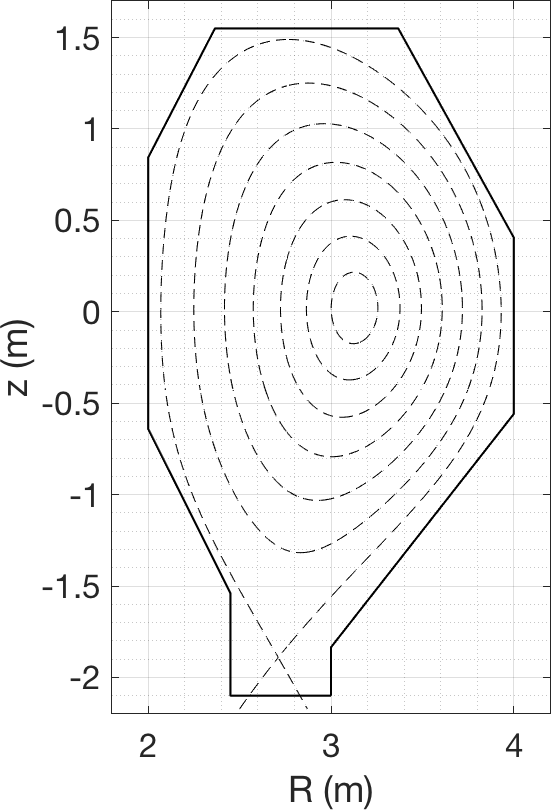}
\hspace{1.5cm}
\includegraphics[height=0.45\textwidth]{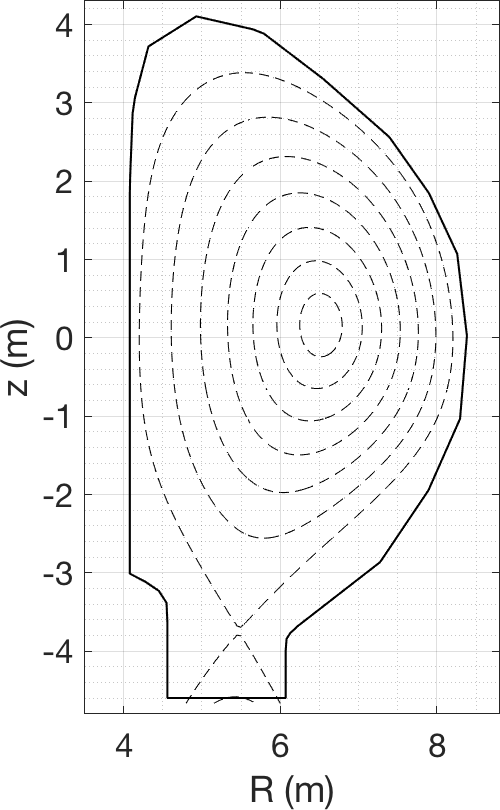}\\
\includegraphics[width=0.48\textwidth]{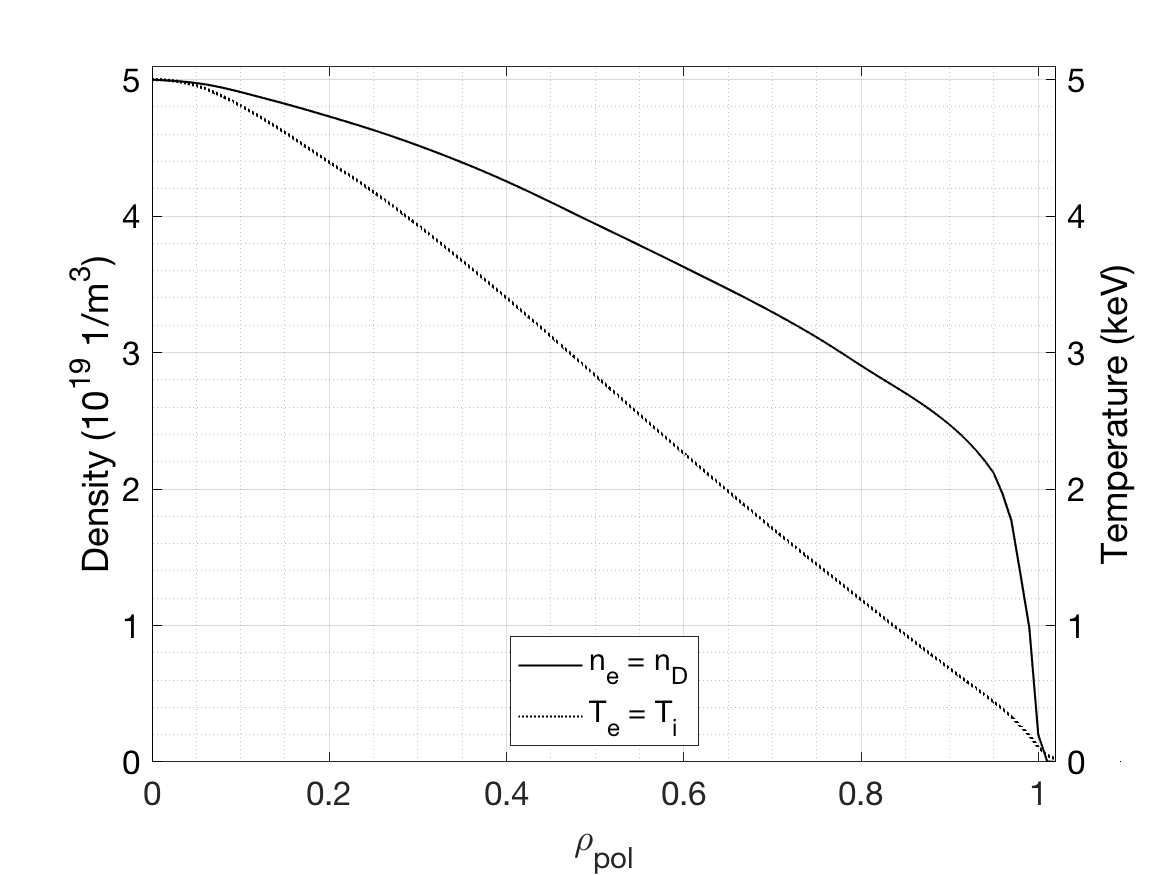}
\includegraphics[width=0.48\textwidth]{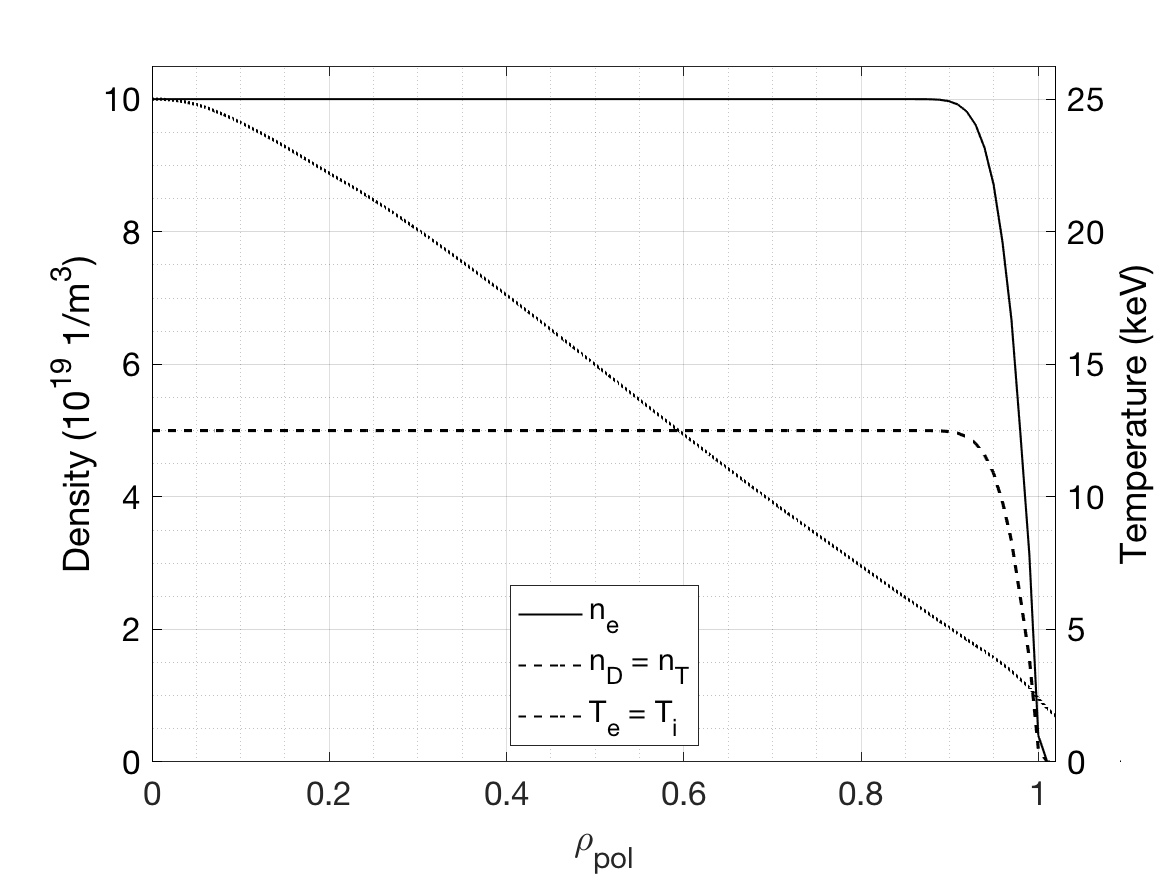}\\
\caption{Equilibrium, wall contour and kinetic profiles for the JET-like case (left) and the ITER-like case (right).}
\label{fig:eq_benchmark}
\end{figure}

The verification with ASCOT4 was performed for two cases representing most common applications for ASCOT: a 2D JET-like case and a 3D ITER-like case. All simulations were performed with guiding-center following with a fixed time step of $10^{-7}$ s. The JET-like case (figure~\ref{fig:eq_benchmark}, left) is based on an axisymmetric analytic equilibrium with 3 T toroidal field and 3 MA current, representing the size and shape of a typical H-mode plasma in the JET tokamak. The equilibrium was converted to a discretized 2D background with a grid resolution of 7-8 mm. A slowing-down simulation of $10^5$ markers representing 120 keV NBI particles was performed both for perpendicular and tangential co-current beam geometries.

The ITER-like (figure~\ref{fig:eq_benchmark}, right) case is similar to the planned ITER baseline 5.6 T / 15 MA plasma scenario. The axisymmetric analytic equilbrium was discretized with a resolution of 5.5 cm and combined with a 3D toroidal ripple field with a resolution of 0.67 degrees. The ripple was calculated using the BioSaw \cite{akaslompolo2015calculating} code. A 3D wall comprised of 86 016 triangles was constructed by revolving a 2D countour around the central axis. For the ITER-like case, $10^5$ isotropic 3.5 MeV fusion alpha particles were simulated, initialized with the AFSI fusion source code \cite{siren2017versatile}.

\begin{figure}[p]
\centering
\includegraphics[width=0.63\textwidth]{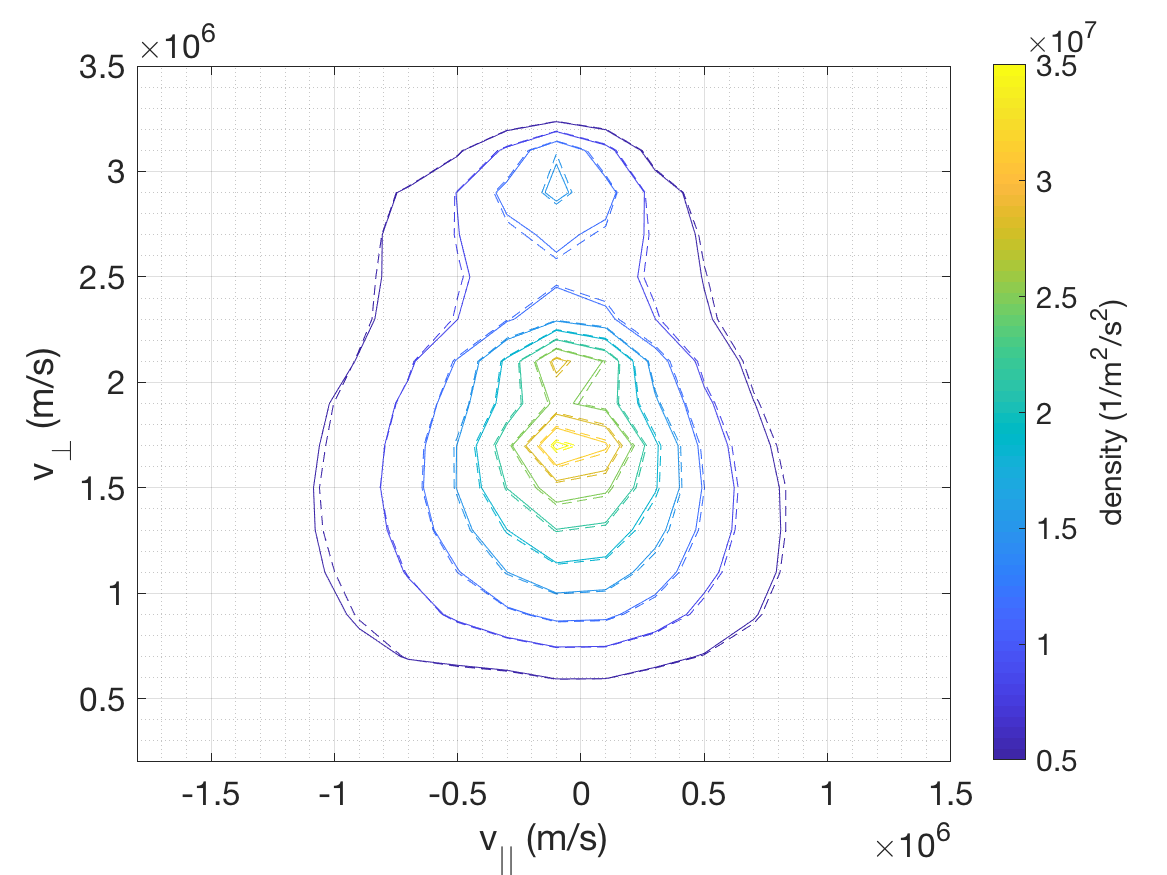} \\
\includegraphics[width=0.63\textwidth]{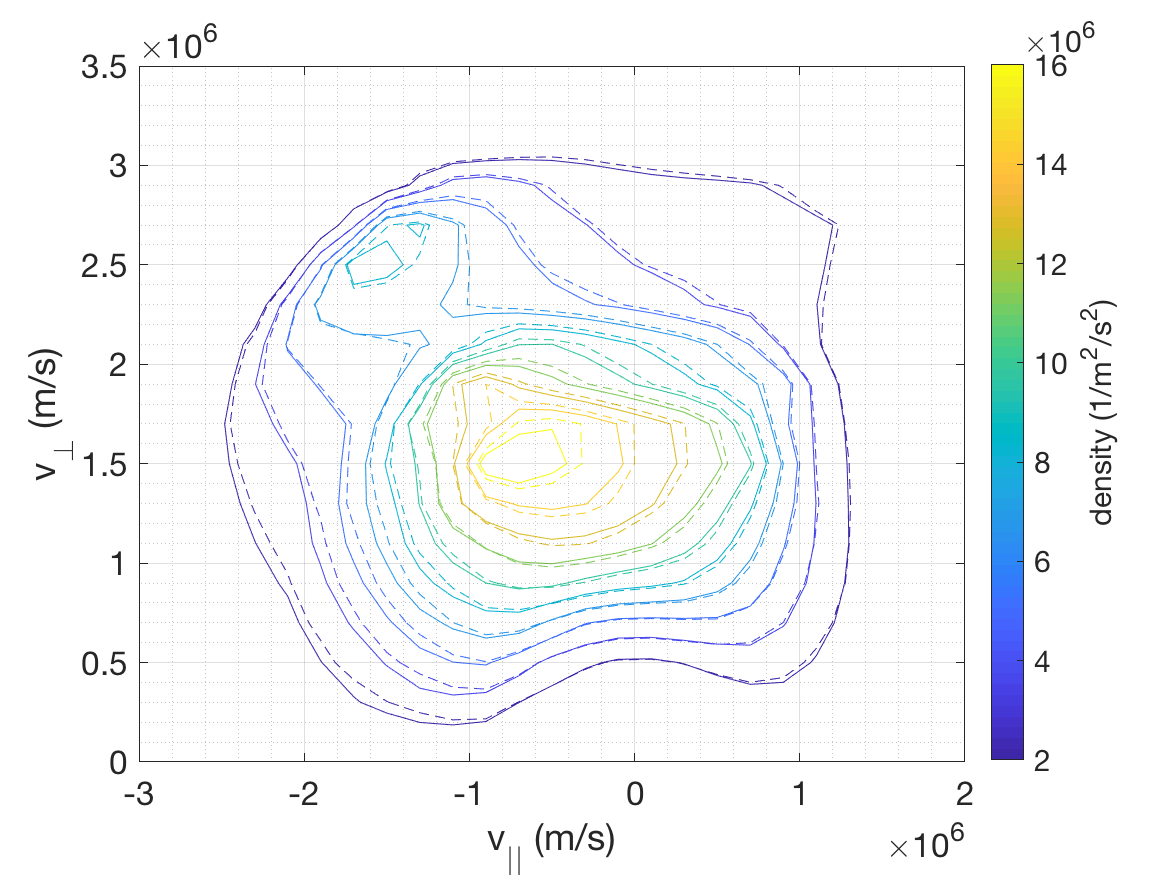} \\
\includegraphics[width=0.63\textwidth]{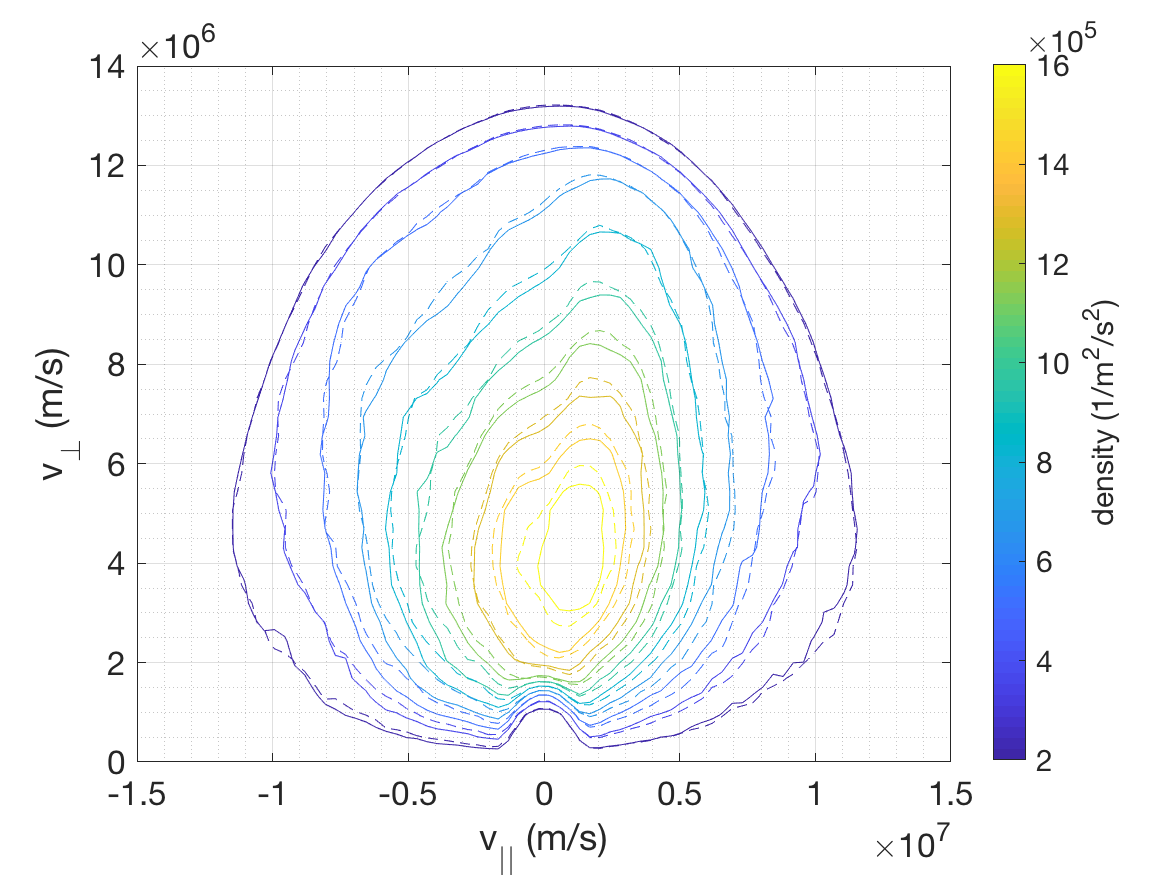}
\caption{Comparison of ASCOT4 (solid) and ASCOT5 (dashed) distribution functions for perpendicular (top) and tangential co-current (middle) 120 keV NBI injected deuterons in a JET-like plasma, and for thermonuclear alpha particles (bottom) in an ITER-like plasma.}
\label{fig:dist_benchmark}
\end{figure}

In both cases, the slowing-down distributions calculated by ASCOT4 and ASCOT5 were found to closely agree both qualitatively and quantitatively (figure~\ref{fig:dist_benchmark}), reproducing the characteristic slowing-down rate and pitch scattering during the slowing-down process. Fast ion losses were likewise in agreement, with total losses reproduced within 1 \% between the two codes.

\section{Performance}

\begin{table}[ht]
\centering
\caption{Comparison of execution times between ASCOT4 and ASCOT5 for the two test cases on different CPU architectures}
\label{table:performance}
\begin{tabular}{l|cc}
& ASCOT4 & ASCOT5 \\
\hline
\emph{JET-like 2D} \\
Skylake 10x48 cores & 0:36:25 & 0:05:24 \\
Knights Landing 10x68 cores & 4:03:50 & 0:24:52 \\
\emph{ITER-like 3D} \\
Skylake 10x48 cores & 5:34:58 & 1:17:00 \\
Knights Landing 20x68 cores & \centering \footnote[1]{Unable to run ASCOT4 due to limited memory} & 2:33:19 
\end{tabular}
\end{table}

The simulations presented in the previous section were repeated on a number of different CPU architectures to evaluate the performance of the code (table~\ref{table:performance}). For the Skylake and Knights Landing CPUs, $N_\mathrm{SIMD} = 8$ markers were simultaneously simulated, corresponding to the 512-bit vector operations (AVX-512). Additionally, on Knights Landing full hyperthreading with 4 threading was used, but hyperthreading was not available on the Skylake system used for the benchmarks.

Initial benchmark results were found to be disappointing, when ASCOT5 was only approximately 2-3 times faster than ASCOT4 in simulations with identical backgrounds, markers and time stepping. Profiling revealed that a significant overhead in ASCOT5 was caused by the random number generation for the collision operator. ASCOT4 uses the standard MT19937 implementation, while in ASCOT5 either standard C library, GSL or MKL random number generators can be used. Even the SIMD-oriented version of the Mersenne Twister was found to significantly increase the execution time due to the memory overhead of the RNG state. Switching to a more memory-efficient RNG algorithm, while ensuring that the simulation results are still converging, mitigated this problem.

On the Skylake system, ASCOT5 was over 6 times faster in the 2D JET-like case, and approximately 4 times faster in the 3D ITER-like case. In the 3D case, cache misses and limited memory bandwidth are likely the reason for not reaching as high an improvement as in the 2D case. On the Knights Landing system the speedup is even greater, as ASCOT5 is able to utilize multiple threads on each core. For the 3D case, it was not possible to run ASCOT4 with 68 processes due to memory limitations. Without shared memory, each ASCOT4 process required approximately 2 GB of memory, exceeding the available 86 GB on the system.

\begin{figure}[ht]
\centering
\includegraphics[width=0.7\textwidth]{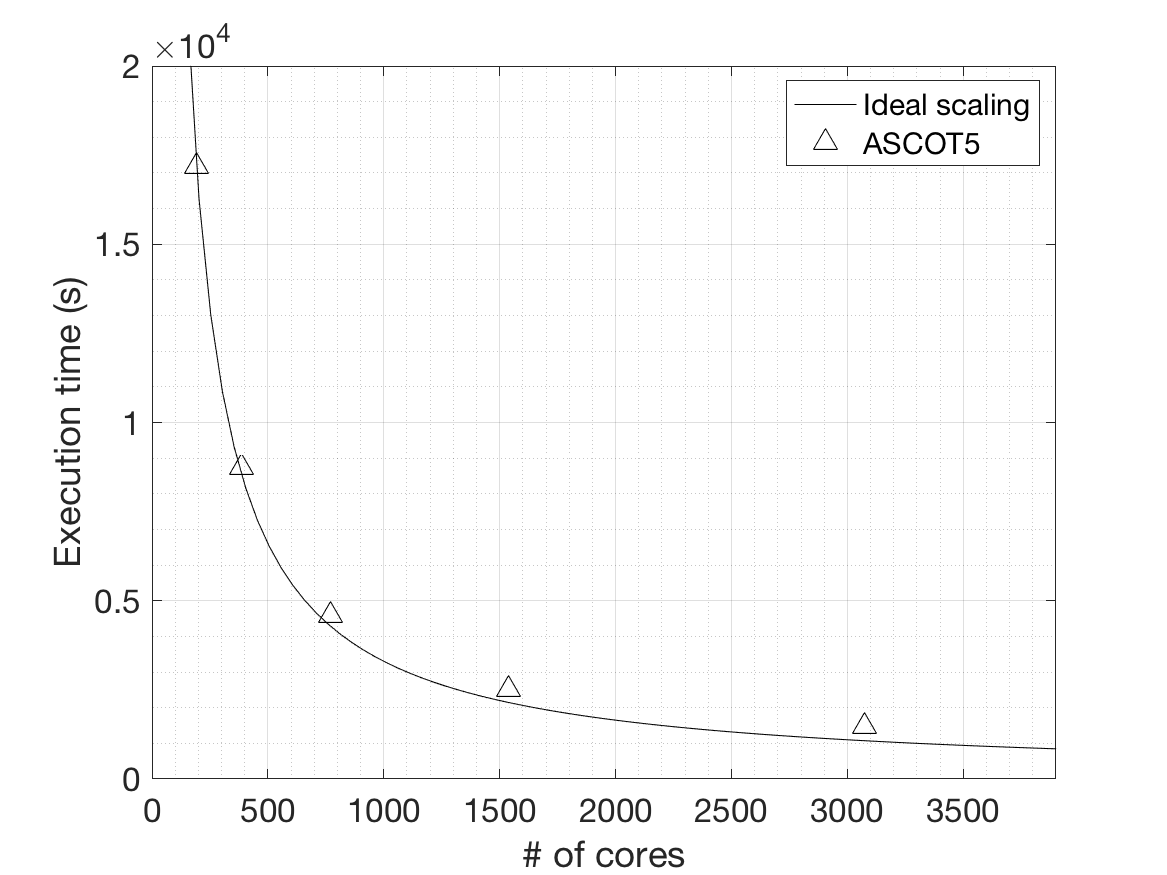}
\caption{Strong scaling of the ASCOT5 code in the ITER-like case using 100k markers. The code scales nearly ideally for typical simulation sizes.}
\label{fig:strong_scaling}
\end{figure}

Finally, a strong scaling test was performed with the ITER-like case. The code demonstrates near ideal scaling, which is to be expected as each marker simulation is independent (figure~\ref{fig:strong_scaling}). Overhead only arises from reading and writing simulation data at the beginning and the end, as well as from underutilization of the SIMD operations at the end of the execution, where the simulation of some markers may take significantly longer than average. Adjusting the number of processes for a sufficiently long simulation time marginalizes the first problem. The latter problem could be mitigated by arranging the order in which the markers are simulated in such a way that the markers that are expected to run longest are simulated first.

\section{Conclusion}

In order to fully utilize modern, highly-parallelized supercomputer hardware, a new multithreaded, SIMD-enabled version of the Monte Carlo orbit-following code ASCOT5 has been written. The code is explicitly parallelized with worker threads that each simulate an array of markers to maximize the utilization of vector instructions. The code is shown to reproduce both the theoretically predicted neoclassical transport as well as simulation results calculated by the previous version of the code for realistic JET- and ITER-like cases.

With the increased performance and memory efficiency of ASCOT5, simulations with nearly an order of magnitude greater marker numbers are now possible with the same computational effort. This allows reaching higher statistical precision for distribution functions as well as for estimates of the for maximum wall loads due to particle losses. Alternatively, faster execution times enable rapid simulations for larger parameter scans for various quantities, and faster turnaround times in coupled transport simulations.

ASCOT5 has been designed from the ground up to be extensible and maintainable with simple interfaces for the different physics modules. Active development is ongoing for additional physics modules, including models for MHD perturbations, time-dependent magnetic and electric fields, turbulent fast ion transport and NBI and fusion particle sources. Planned performance improvements include optimizing the data structures for specific simulation geometries to improve memory efficiency. Finally, the OpenMP-based parallelization and offloading techniques are extensible to other architectures such as GPUs. This will enable ASCOT to flexibly utilize a wide variety of computing platforms using the same unaltered codebase.

\bibliographystyle{unsrt}
\bibliography{main}

\end{document}